# Singular enhancement of emission of entangled photons for surface plasmon polaritons

V. Hizhnyakov

Institute of Physics, University of Tartu, Riia 142, 51014 Tartu, Estonia
E-mail: hizh@fi.tartu.ee

**Abstract.** Emission of photon pairs by metal-dielectric interface in case of a superluminally propagating laser-induced nonlinear polarization is considered. Using oblique incidence of the laser wave it is possible to tune frequencies of generated photons due to the Doppler effect. For every frequency smaller than frequency of induced polarization there exist an excitation angle when singular enhancement of the emission takes place. Emitted photon pairs are in the polarization entangled state. The yield of emission is estimated.

1. Introduction

In this communication we consider the laser-induced emission of entangled photon pairs by a metal -dielectric interface. The process takes place due to the nonlinear interaction of the laser light with surface plasmon polaritons (SPPs). The characteristic feature of this process is a possibility to introduce a moving reference frame in which the phase of the laser-induced nonlinear polarization depends only on time. The latter polarization results in the periodical oscillations in time of the optical length of the interface. Therefore the case under consideration is closely related to the dynamical Casimir effect – generation of photon pairs in resonator with periodically changing in time its length (see, e.g. [1-12]). This emission has also some analogy with the usual spontaneous down conversion [1], [2] - generation of pairs of photons in a bulk in expense of propagating photons of a laser.

To observe the dynamical Casimir effect, one needs to move the plates of the resonator with the velocity comparable to velocity of light [3]. This condition is difficult to fulfill. Therefore the intensity of corresponding emission is usually extremely weak. But if to use a superconducting circuit consisting of a coplanar transmission line then one can achieve for microwaves rather strong changing in time of the electrical length and to strongly enhance the intensity of emission. Recently in this way the dynamical Casimir effect was observed in 10 GHz diapason [4].

In visible the dynamical Casimir effect may be also enhanced if to use strong lasers excitation allowing to remarkably change in time the optical length of a dielectric [5-9]. Moreover, as it was shown in Refs, [6,7] there should exist a characteristic, although usually rather large intensity of the laser light when the two-photon emission is strongly enhanced. This occurs if the amplitude of oscillations of the optical length coincides with the wavelength of the excitation. Below we will show that this enhancement can be especially strong in case of metal-dielectric interface if to use oblique incidence of the laser wave.

## 2. Quantum emission in case of oscillating optical length

Let us consider dynamical Casimir effect in case of steadily oscillating optical length (eikonal) of a resonator. We suppose that one of the mirrors of the resonator is situated at coordinate $x = 0$ and another oscillates near coordinate $x = L$. The optical length of the resonator equals

$$L_t = L + a\cos(\omega_0 t) \qquad (1)$$

Here $a$ and $\omega_0$ are the amplitude and the frequency of the oscillations, respectively. If the reason of oscillations is the nonlinear interaction of the placed in resonator medium with the laser light (see Fig. 1) then

$$a = 2\pi l_0 \chi^{(m)} E_0^{m-1} / (m-1) \qquad (2)$$

and $\omega_0 = (m-1)\omega_L$, where $\omega_L$ is the frequency of the laser light, $E_0$ is the amplitude of its electric field, $l_0$ is the length of the excited area, $m$ is the order of the working nonlinearity (we consider only nonlinearity of second ($m = 2$) and third ($m = 3$) order). In case of emission caused by the second order nonlinearity $\omega_0 = \omega_L$ and $a \propto \sqrt{I_0}$ while in case of the emission caused by the third order nonlinearity $\omega_0 = 2\omega_L$, $a \propto I_0$, where $I_0 = E_0^2 / Z_0$ is the intensity of the laser excitation, $Z_0$ is the impedance of free space.

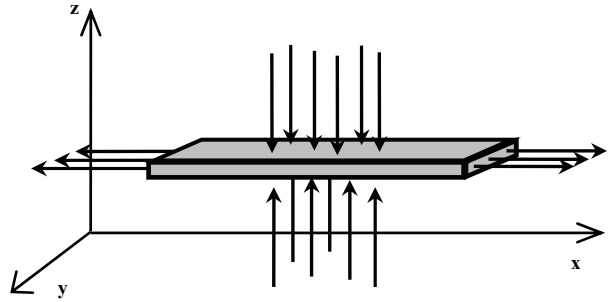

Fig. 1. Scheme of spontaneous quantum emission of dielectric exposed in a standing laser wave.

To find the number of generated photons due to the dynamical Casimir effect one usually calculates the Bogoliubov transformation of the field operators $\hat{b}_k^{(1)} = \mu_k \hat{b}_k + \nu_k \hat{b}_k^+$ caused by the changing in time of the optical length [5]. Here $\hat{b}_k$ and $\hat{b}_k^{(1)}$ are the annihilation operators of the mode $k$ at small and large time, respectively, $\mu_k$ and $\nu_k$ are the complex parameters satisfying the condition $|\mu_k|^2 = 1 + |\nu_k|^2$. The number of generated photons then equals $N_k = |\nu_k|^2$. Here we use another, although similar way to find $N_k$ based on calculation of the large time asymptotic of the pair correlation function of the field operator $\hat{A}_k(t) = \sqrt{\hbar/2\omega_k}\, \hat{b}_k^{(1)} e^{-i\omega_k t}$ [6]. In the large $t$ limit this function equals

$$d_k(\tau) = \langle 0 | \hat{A}_k^+(t+\tau)\hat{A}_k(t) | 0 \rangle = (\hbar/2\omega_k) N_k e^{i\omega_k \tau} \qquad (3)$$

where $|0\rangle$ is the initial zero-point state.

The field operator in the resonator satisfies the wave equation in vacuum and the boundary conditions $\hat{A}(x=0,t) = \hat{A}(x=L_t,t) = 0$. Therefore the $x$ dependence of the field operator $\hat{A}$ can be taken as a linear combination of the operators $\hat{A}_k \sin(\pi k x/L_t)$. Taking $\hat{A} = \sum_k \hat{A}_k \sin(\pi k x/L_t)$ and inserting this operator to the wave equation we get in the large $L$ limit

$$\sum_{k=-\infty}^{\infty}\left[\left(\ddot{\hat{A}}_k + \omega_k^2 \hat{A}_k\right)\sin\left(\pi k x L_t^{-1}\right) - k\pi x L_t^{-2}\left(2\dot{L}_t\dot{\hat{A}}_{k,\bar{q}} + \ddot{L}_t \hat{A}_k\right)\cos\left(\pi k x L_t^{-1}\right)\right] = 0 \qquad (4)$$

We use now the equation for the Fourier series of the saw-tooth wave

$$x = -\sum_{j\neq 0}(-1)^j j^{-1}\sin(jx), \quad x - \pi \in 2\pi n$$

where $n$ is the natural number, which gives for $\hat{A}_k = (-1)^k \hat{A}_k$ the equation

$$\ddot{\hat{A}}_k + \omega_k^2 \hat{A}_k = \omega_k \hat{B}_k \qquad (5)$$

where $\omega_k = \pi c_0 k/L$, $\hat{B}_k = \sum_{j\neq k} j\left(2\dot{L}_t\dot{\hat{A}}_j + \ddot{L}_t \hat{A}_j\right)/\pi c_0(j^2 - k^2)$, $c_0$ is the light velocity in vacuum. Eq. (5) can be presented in the integral form

$$\hat{A}_k(t) = \hat{A}_k^{(0)}(t) - \int_0^t \sin(\omega_k(t-t'))\hat{B}_k(t')dt' \qquad (6)$$

where $\hat{A}_k^{(0)}(t)$ is the operator of the undisturbed field. In case of periodically time-dependent $L_t$ one may consider the $t \to \infty$ limit. Then nonzero contributions to integral in Eq.(6) come from the terms $\propto e^{\pm i(\omega_0 - \omega_k - \omega_j)t}$ with $\omega_j + \omega_k \cong \omega_0$. Therefore $2\dot{L}_t\dot{\hat{A}}_j + \ddot{L}_t \hat{A}_j \cong a(\pi c_0/L)^2(j^2 - k^2)\cos(\omega_0 t)\hat{A}_j$. As a result the factor $j^2 - k^2$ in the equation for $\hat{B}_k$ cancels and we get

$$\hat{B}_k = 2a\cos(\omega_0 t)\hat{Q}, \quad \hat{Q} = L^{-1}\sum_{j=1}^{\infty}\omega_j \hat{A}_j(t) \qquad (7)$$

Inserting Eqs. (6) and (7) into Eq. (3) and taking into account that the negative frequency term $\propto e^{i\omega_k \tau}$ in the correlation function $d_k(\tau)$ comes from $\sin(\omega_k(t+\tau-t'))$ in Eq. (6) we get

$$N_k(t) \simeq \frac{a^2 \omega_k}{2\hbar L}\int_0^t\int_0^t dt_1 dt_1' e^{i(\omega_0 - \omega_k)(t_1 - t_1')} D(t_1, t_1') \qquad (8)$$

Here $D(t,t') = L\langle 0|\hat{Q}^+(t)\hat{Q}(t')|0\rangle$. Taking now into account that in the $t \to \infty$ limit the pair-correlation function depends on time difference ($D(t,t') = D(t-t')$) we get the following equation for the rate of creation of photons with the frequency $\omega_k$:

$$\dot{N}_k = \left(a^2/2\hbar L\right)\omega_k D(\omega_0 - \omega_k) \qquad (9)$$

where $D(\omega)$ is the Fourier transform of $D(t)$. In the $L \to \infty$ limit the discrete quantities $k$ and $\omega_k$ can be replaced by the continuous variables $k$ and $\omega$. This gives the following equation for the spectral density of the emission rate:

$$I(\omega) \equiv dN(\omega)/dt = \left(a^2/2\pi c_0 \hbar\right)\omega D(\omega_0 - \omega)\Theta(\omega_0 - \omega) \qquad (10)$$

where $\Theta(\omega)$ is the Heaviside step function. If the amplitude of oscillations of the optical length $a$ is small as compared to the wave length $\lambda_\omega$ of emitted photons then $\hat{Q}^{(0)}(t) = L^{-1}\sum_{k=1}^{\infty}\omega_k \hat{A}_k^{(0)}(t)$ and $D(\omega) \approx D^{(0)}(\omega) = \hbar\omega/c_0$. This gives [6,7]

$$I^{(0)}(\omega) \approx (v^2/2\pi)\Omega(1-\Omega)\Theta(1-\Omega) \quad (11)$$

where $\Omega = \omega/\omega_0$ is the dimensionless frequency of emission, $v = a\omega_0/c_0$ is the dimensionless velocity of oscillations of optical length which is here much less than unity. According to Eq. (11) the spectrum of two-photon emission under consideration is continuous. Analogous result was obtained theoretically and verified experimentally in [4].

If $v$ is not small then

$$\hat{Q}(t) = \hat{Q}^{(0)}(t) - 2aL^{-1}\int_{-\infty}^{t} dt' \sum_k \omega_k \sin(\omega_k(t-t'))\cos(\omega_0 t')\hat{Q}(t') \quad (12)$$

In the $t \to \infty$ limit Eq. (12) can be easily solved for the Fourier transforms of the operators. One gets

$$\hat{Q}(\omega) = \hat{Q}^{(0)}(\omega)/\left(1 - v^2 G(\omega/\omega_0)G^*(1-\omega/\omega_0)\right) \quad (13)$$

where $G(x) = \left[2 + x\ln((1-x)/(1-x))\right]/2\pi + i(x/2)\Theta(1-x)$. This gives [8,9]

$$I(\omega) = I^{(0)}(\omega)/\left|1 - v^2 G^*(\Omega)G(1-\Omega)\right|^2 \quad (14)$$

Note that $I(\omega_0 - \omega) = I(\omega)$. This relation reflect the fact that the emission occurs by pairs of photons with the frequencies $\omega$ and $\omega_0 - \omega$. Both photons are emitted in the same direction and propagate together. This important property of the two-photon emission under consideration follows from the fact that Eq. (14) holds also for the case of single oscillating mirror: both photons are created on the same side of the mirror. Presented above consideration holds for 1D case. In case of emitting ribbon one should multiply Eq.(14) by the numbers of emitting modes with the frequencies $\omega$ and $\omega_0 - \omega$.

According to Eq. (14) the emission caused by the dynamical Casimir effect is determined by the dimensionless velocity $v$ of the oscillation of the optical length. If $v \ll 1$ then the denominator in Eq. (14) is close to unity and the intensity of emission increases with $v$ as $v^2$. However if $v \gg 1$ then the denominator increases with $v$ as $v^4$ and the intensity decreases with $v$ as $v^{-2}$. The crossover from increasing to decreasing takes place for $v = v_r = 1/|G(1/2)| \approx 2.94$. For this $v$ the resolvent in Eq. (14) diverges at $\omega = \omega_0/2$ and the emission is resonantly enhanced. Note that $v_r$ is close to $\pi$ when $2a = \lambda_{\omega_0}$. Consequently the resonance corresponds to the coincidence of the full amplitude of oscillations of the optical length with the wave-length of excitation.

### 3. Superluminal propagation of nonlinear polarization

Let us consider two dielectrics with refractive indexes $n_0$ and $n_1$ separated by thin metallic film. In the interface of this structure can exist electromagnetic surface waves, - SPPs. We suppose that the interface is excited by strong monochromatic plane wave of a laser from the side of a dielectric with refractive index $n_0$. The induced by laser electric field in the coordinate $x$ of the interface at time moment $t$ reads

$$E(t,x) = E_0 \cos\left(\omega_L\left(t - \sin\varphi n_0 x/c_0\right)\right) \quad (15)$$

where $\omega_L$ is the frequency of the laser wave, $\varphi$ is the angle between the direction of excitation and the normal to the interface (see Fig. 2). This field nonlinearly interacts with surface. The effect of this interaction to a component of the vector potential $A \propto e^{i\omega t}$ of SPP with the frequency $\omega$ is described by the nonlinear wave equation

$$\left(\partial^2/\partial t^2 - c^2 \partial^2/\partial x^2\right) A = 4\pi\omega^2 p_m(t,x) A \qquad (16)$$

where $c = c_0/n$ is the phase velocity and $n \equiv n_\omega$ is the refractive index of the SPP mode under consideration, $p_m(t,x)$ is the nonlinear polarizability of the $m$'th order. We will consider the nonlinear effects of the second- and third-order described by $p_2(t,x) = \chi^{(2)} E(t,x)$ and $p_3(t,x) = \chi^{(3)} E^2(t,x)$, respectively, where $\chi^{(2)}$ and $\chi^{(3)}$ are the second- and the third-order nonlinear susceptibilities of the interface. We take into account that the laser field $E(t,x)$ can be presented in the form $E(t,x) = E_0 \cos\left(\omega_L \left(t - \alpha x/c\right)\right)$, where $\alpha = \sin(\varphi) n_0/n$.

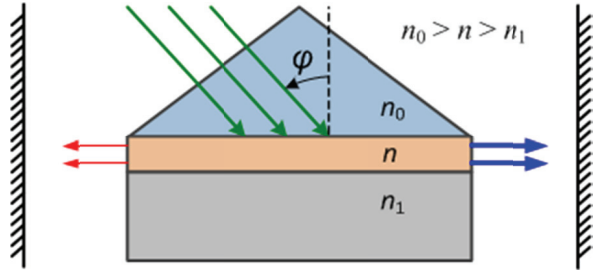

Fig. 2 Scheme of excitation and emission of surface plasmon polaritons (SPPs) in the metallic interface; φ is the excitation angle, $n_0$, and $n_1$ are refractive indexes of dielectrics, n is the refractive index of SPP.

In the case under consideration the time dependent part of the nonlinear polarizability $p_m(t) - p_m(0) \propto \cos\left((m-1)\omega_0(t - \alpha x/c)\right)$ moves along the interface with the velocity $c/\alpha$ and oscillates in time with the frequency $\omega_0 = (m-1)\omega_L$. If $n_0 > n$ then one can get any value of $\alpha$ between 0 and 1 with proper choice of $\varphi \leq \varphi_0 = \arcsin(n/n_0)$. If $\alpha = 1$ then $\varphi = \varphi_0$ and the velocity is the same as the phase velocity of the mode under consideration; this case corresponds to the phase matching of the nonlinear polarization and of the SPP mode.

We consider excitation with $\varphi < \varphi_0$. Then $\alpha < 1$ and the induced polarization propagates along the interface superluminally, i.e. faster than SPP under consideration. In this case one can introduce the moving with the velocity $\upsilon = \alpha c < c$ reference frame

$$x' = \gamma(x - \upsilon t), \quad t' = \gamma\left(t - x\upsilon/c^2\right) \qquad (17)$$

($\gamma = 1/\sqrt{1-\alpha^2}$, $\alpha = \upsilon/c$) in which for given time $t'$ the field is the same for all $x'$. (Moving reference frames for description of nonlinear optical effects was earlier used in [17-19]). The nonlinear wave equation in this reference frame gets the form

$$\left(\partial^2/\partial t'^2 - c^2 \partial^2/\partial x'^2\right) A' = 4\pi\omega'^2 p'_m(t) A' \qquad (18)$$

where $A' \propto e^{i\omega' t'}$ depends only on time, $\omega' = \omega\sqrt{(1-\alpha)/(1+\alpha)}$ is the frequency of the mode and

$$p'_m(t) = \frac{\chi^{(m)} E_0^{m-1}(1+\alpha)}{(1-\alpha)(m-1)}\left(m - 2 + \cos(\omega'_0 t')\right) \qquad (19)$$

is the nonlinear polarizability in the moving reference frame, $\omega'_0 = \omega_0\sqrt{1-\alpha^2}$. Replacing in Eq. (18) $\omega'^2$ by $-\partial^2/\partial t'^2$ we get the wave equation with the time-dependent refractive index $n'_{t'}$:

$$\left(\partial^2/\partial t'^2 - (c_0^2/n_{t'}'^2)\partial^2/\partial x'^2\right)A' = 0 \qquad (20)$$

Here $n_{t'}' = n\sqrt{1+4\pi p_m'(t')} \cong n(1+2\pi p_m'(t'))$ (we consider the case of relatively small changing of $n$ in time). Consequently the laser excitation of the interface of a slab leads in the moving reference frame to periodical in time changing of the optical length of the slab.

## 4. Singular enhancement of two-photon emission

Let us place the slab into a resonator with mirrors moving together with the reference frame. In this reference frame the slab changes (linearly) its coordinates in time. However its optical length does not depend on its position. Therefore the change in time of the optical length of the interface in the slab leads to the same changes in time of the optical length of the entire resonator as the slab would be in the rest in the moving reference frame. Therefore the field operator in the resonator outside the slab satisfies the wave equation in vacuum and the boundary conditions $\hat{A}(x'=0,t) = \hat{A}(x'=-L_t',t) = 0$ where $L_t' = \gamma(L + a\cos(\omega_0' t'))$, $L = L_0 + a(m-2) + l_0(n-1)$, $a$ is the amplitude of oscillations of the optical length and $l_0$ is the length of the excited interface in the laboratory reference frame. The derived above equation (14) holds also in the case under consideration but in the moving reference.

In the moving reference frame time and length are $\sqrt{1-\alpha^2}$ times shorter than in the laboratory reference frame. Therefore $dt' = \sqrt{1-\alpha^2}\,dt$ and

$$v' = v\sqrt{(1+\alpha)^3/(1-\alpha)} \qquad (21)$$

Besides, due to the Doppler effect emissions forward and backward have different frequencies: $\omega = \omega'\sqrt{(1\pm\alpha)/(1\mp\alpha)}$ (here and below the upper sign correspond to emission forward, while the lover sign to emission backward). This gives $d\omega' dt' = (1\mp\alpha)d\omega dt$. As a result we get the following spectral density rate for two-photon emission forward:

$$I(\omega) \approx \frac{v'^2(1-\alpha)\Omega(1-\Omega/(1+\alpha))\Theta(1-\Omega/(1+\alpha))}{2\pi(1+\alpha)\left|1-v'^2 G^*(\Omega/(1+\alpha))G(1-\Omega/(1+\alpha))\right|^2} \qquad (22)$$

In case of emission backward one should replace $1-\alpha$ by $1+\alpha$ and $\Omega/(1+\alpha)$ by $\Omega/(1-\alpha)$. Note that total number of photons emitted forward and backward is the same and in both directions photons propagate by pairs.

According to Eq. (22) intensity $I(\omega)$ is a singularly enhanced at frequency $\omega_s = \omega_0(1+\alpha_s)/2$ and angle $\varphi_s = \arcsin(\alpha_s n_{\omega_s}/n_0)$. Here $\alpha_s$ is determined by the equation $v^2|G(1/2)|^2(1+\alpha_s)^3 + \alpha_s = 1$ (giving $\alpha_s = -1 + \sqrt[3]{\sqrt{p+p^3/27}+p} - \sqrt[3]{\sqrt{p+p^3/27}-p}$, where $p = 1/|G(1/2)|^2 v^2$). Indeed for $\alpha \to \alpha_s$ and $\Omega \to (1+\alpha_s)/2$ the denominator in Eq. (22) tends to zero giving a sharp peak in the spectrum (see Figs. 3; in our numerical calculations we used finite spectral resolution $\Gamma = 10^{-3}\omega_0$).

The total intensity of emission $J = \int_0^{\omega_0(1+\alpha)} d\omega \dot{N}(\omega)$ is also singularly enhanced (see Fig.4). If $n_0 > n$ then for every $v$ there is an angle $\varphi$ for which singular enhancement of the emission takes place. The reason of the enhancement is the coincidence of the amplitude of oscillations of the optical length with the wavelength of excitation in the moving reference frame. Both, frequency of the peak and the value of $\varphi$ monotonously diminish with increasing of the excitation strength (see Fig. 5).

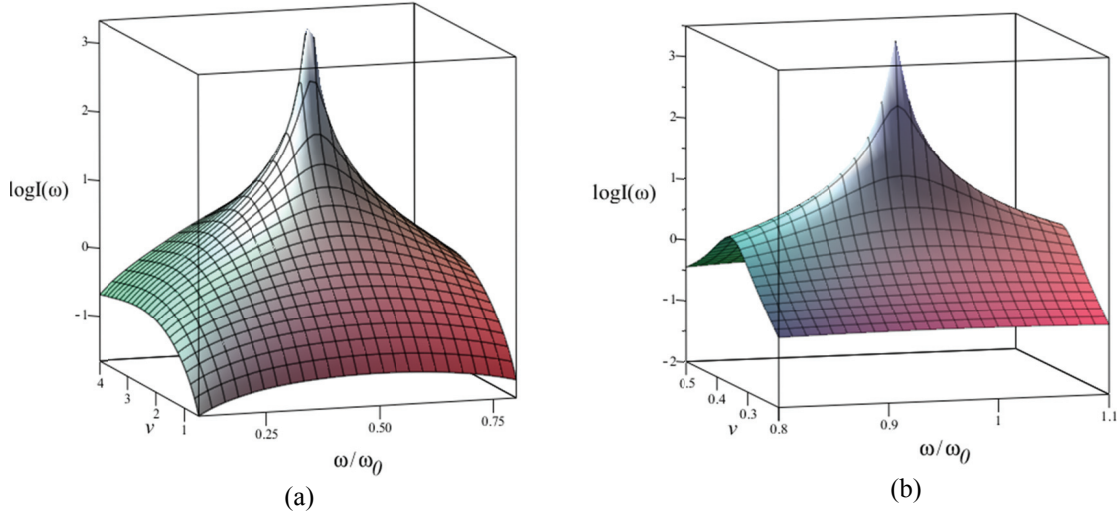

(a)  (b)

Fig. 3 Dependence of emission rate on the frequency $\omega/\omega_0$ and velocity of oscillations of the optical length $v \propto \sqrt{I_0^{m-1}}$ for (a) $\sin\varphi = 0.5\, n/n_0$; (b) $\sin\varphi = 0.9\, n/n_0$. $I_0$ is the intensity of the laser excitation: ($m$ is the order of the nonlinearity).

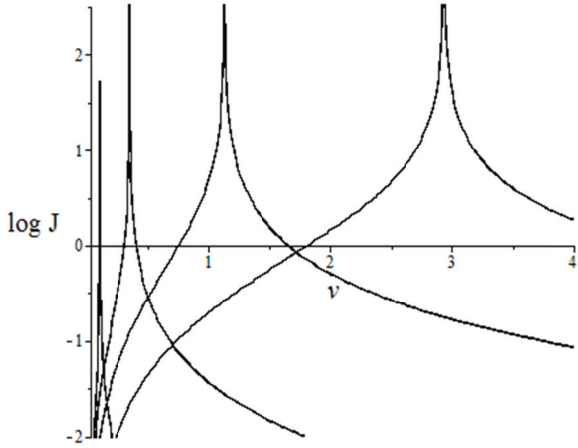
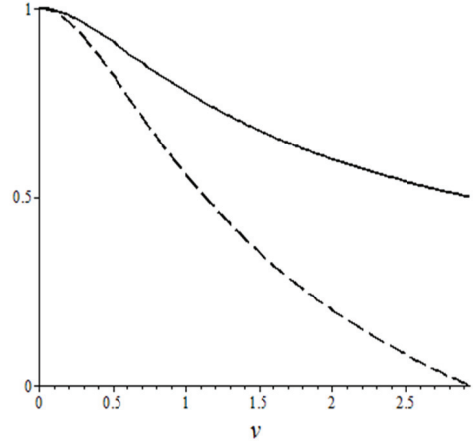

Fig. 4 Dependence of the integral intensity of emission on $v \propto \sqrt{I_0^{m-1}}$ for direction parameters $\alpha = 0.99, 0.9, 0.5$ and $0.1$ (from left to right).

Fig. 5 Dependence of $\alpha = \sin(\varphi) n_0/n$ (dash line) and dimensionless frequency $\omega/\omega_0$ (solid line) of the emission peak on $v \propto \sqrt{I_0^{m-1}}$.

In case of $v \ll 1$ (weak excitation) $\alpha_s \approx 1 - 0.926 v^2$ and $\omega \approx \omega_0 (1 - 0.463 v^2)$. In this case the frequency of the emission peak is close to $\omega_L$ (for second-order nonlinearity) or to $2\omega_L$ (for third-order nonlinearity). At that the singular enhancement takes place at the angle close to the Kretschmann configuration $\varphi = \arcsin(n_{\omega_0}/n_0)$.

Presented above consideration holds for 1D case. To find the number of emitted photons by an excited ribbon of the interface one should multiply Eq. (22) by the numbers of emitting modes with the frequencies $\omega$ and $\omega_0(1 \pm \alpha) - \omega$. These numbers are approximately equal $d_0/\lambda_\omega$ and $d_0/\lambda_{\omega_0(1\pm\alpha)-\omega}$, supposing that $d_0 \ll l_0$ (here $d_0$ is the width of the excited ribbon). As a result the spectral intensity of emission forward by the interface ribbon gets the form

$$I(\omega) \approx \frac{v'^2(1-\alpha)d_0^2\Omega^2(1+\alpha-\Omega)^2\Theta(1+\alpha-\Omega)}{2\pi\lambda_{\omega_0}^2\left|1-v'^2 G^*(\Omega/(1+\alpha))G(1-\Omega/(1+\alpha))\right|^2} \qquad (23)$$

## 5. Polarization entanglement of emission

Let us chose symmetry axes so that the normal to the interface is oriented in $z$ direction and consider the polarization of the plane wave of excitation in $y$ direction. In this case the mirror symmetry in $z$ direction is absent. Therefore the components $\chi^{(2)}_{y,yz}$ and $\chi^{(2)}_{y,zy}$ of the second-order nonlinear susceptibility tensor describing two-photon emission in $x$ direction differ from zero (here the first index corresponds to polarization of excitation, while last two indexes correspond to the polarization of emitted quanta). This means that one of photons in the emitted pair of photons has $y$ polarization and another has $z$ polarization. Due to identity of both emitted photons their wave function must be the linear combination of the wave functions with both polarizations

$$\left|\psi\right\rangle_{xz} = \frac{1}{2}\left(\left|\psi_{1x}\right\rangle\left|\psi_{2y}\right\rangle + \left|\psi_{2x}\right\rangle\left|\psi_{1y}\right\rangle\right) \qquad (24)$$

where $\left|\psi_{n\alpha_0}\right\rangle$ is the wave function of a photon number $n=1,2$ with polarization index $\alpha_0$. This wave function of the pair of emitted photons describes the basic for the quantum informatics Bell-type state [7]. This state cannot be presented as a product of two one-photon states (otherwise there will be a possibility to find both photons with the same polarization which is impossible here). This means that the emitted photons are in the polarization entangled state. If to divide photons directionally and to choose the polarization of one of them (signal) to be, e.g. $x$, then polarization of another photon (idler) will be $y$. However if to choose polarization of the signal photon to be $y$ then the idler photon will be polarized in $x$ direction. It is just what one needs for using polarization entangled photons for quantum cryptography.

## 6. Discussion

To estimate the intensity of the two-photon emission under consideration in addition to studied above enhancement one should take into account SPP-induced enhancement of the emission ($\eta_e$) and excitation (laser, $\eta_L$) fields. This results in $\eta_L \eta_e^2$ times enhancement of the nonlinear susceptibility $\chi^{(2)}$. In case of gold or silver films $\eta_e$ may exceed 10. E.g. using 60 nm silver film on a dielectric with the refractive index $n=1.5$ one gets for emission with the wave length $\lambda_\omega = 1000$ nm $\eta_e \approx 35$ [23]. (Note that for $n=2.5$ and $\lambda_\omega = 1000$ nm one gets $\eta_e \approx 55$ [23].) We consider the case when the excitation angle differs from that of the Kretschmann configuration for the laser frequency $\omega_L$. This excitation does not create SPP with frequency $\omega_L$. Therefore the field of excitation is not enhanced and $\eta_L \sim 1$. The dimensionless velocity of oscillations of the optical length can be now estimated as

$$v \sim 10^4 \chi_0^{(2)} I_0^{1/2} l_0 / \lambda_0 \qquad (25)$$

Here $\chi_0^{(2)} \sim 10^{-11}$ m/V, which is a typical non-enhanced value of $\chi^{(2)}$ for dielectrics. Taking $l_0/\lambda_0 \sim 10^3$ one finds $v \sim 3$ for $I_0 \sim 10^8$ W/m². For the laser light focused to $\sim 10^{-5}$ m² area this corresponds to the pulse of moderate power 1000 W. Such a pulse with picosecond duration has $\sim 10^{10}$ photons with the total energy $\sim 10^{-9}$ J. If to take the width of the excited ribbon $d_0 \sim 10^2 \lambda_0$ then the total number $N$ of generated photons is $\sim 10^6$, which corresponds to the yield $\sim 10^{-4}$ photon pairs per every photon of excitation. It is much bigger than the yield $10^{-15}$, which one gets in case spontaneous down conversion in dielectrics [8].

As it was shown in [9], SPP-supported nonlinear optical processes can be strongly enhanced by surface nanostructuring. Using this method one can enhance also the considered here two-photon emission. Indeed, a grating of the metallic film of the interface with the period $a$ brings an additional wave number $k_a = 2\pi/a$ to the problem. The reflection of SPP with the frequency $\omega_L$ and the wave number $k_L$ on this grating results in appearance of SPPs with the same frequency but with the wave numbers $k_{nL} = k_L - nk_a$, where $n = \pm 1, \pm 2, ...$ For properly chosen $a$ these SPPs (e.g. the one with the wave number $k_{1L}$) can be excited at considered here oblige excitation. This will result in strong enhancement of the superluminally moving polarization allowing one to get $\eta_L \gg 1$. This make possible to increase additionally the yield of the two-photon emission under consideration more than two orders of magnitude.

The third order nonlinearity described by the susceptibility $\chi^{(3)}$ causes emission of photon pairs with higher frequency (in this case $\omega_0 = 2\omega_L$) and smaller intensity. To estimate the yield of this emission we take the coefficient of the intensity-dependent refractive index in the metal-dielectric interface $n_2 \sim 10^{-14}$ m²/W [9]. (This is the value of $n_2$ for gold-dielectric interface. Standard value of $n_2$ for dielectrics is $\sim 10^{-19}$ m²/W. Consequently the enhancement factor of $\chi^{(3)}$ by SPPs with gold film is $\eta_L^2 \eta_e^2 \sim 10^5$. For SPPs with silver film one can get two orders of magnitude larger enhancement of $\chi^{(3)}$.) If $\lambda_0/l_0 \sim 10^{-3}$ then the dimensionless velocity $v \sim 3$ will be achieved for intensity $I_0 \sim 10^{12}$ W/m². This intensity can be in addition reduced more than two orders of magnitude if to use surface nanostructuring.

In conclusion, the new method of generation of polarization entangled photons is proposed. According to presented estimations the method is more efficient than the standard method based on the spontaneous down conversion in dielectrics.

**Acknowledgement**

The research was supported by Estonian research projects SF0180013s07, IUT2-27 and by the European Union through the European Regional Development Fund (project 3.2.0101.11-0029).